\documentclass[16pt,a4]{article}
\usepackage{epsfig}

\title{Film Thickness Controlled Photo-luminescence Emission in ZnO:Si 
Nano-composites}

\author{Shabnam$^a$, Chhaya Ravi Kant$^b$ and P.
Arun$^c$\footnote{email:arunp92@physics.du.ac.in, Telephone:091 011
29258401, Fax: 091 011 27666220} \\ 
\\
$^a$ University School of Basic and Applied Sciences,\\
Guru Gobind Singh Indraprastha University,\\ New Delhi, India.\\
\\
$^b$Department of Applied Sciences,\\ 
Indira Gandhi Institute of Technology,\\ 
Guru Gobind Singh Indraprastha University,\\ Delhi 110 006, India.\\
\\
$^c$Material Science Research Lab,\\ S.G.T.B. Khalsa College,\\
University of Delhi, Delhi - 110 007, India\\
}

\begin{document}
\maketitle

\begin{abstract}
As grown ZnO:Si nanocomposite films fabricated by thermal evaporation showed 
broad photoluiminescence (PL) spectra due to merging of four prominent peaks 
viz 370, 410, 470 and 520 nm. Investigations revealed the role of interface in 
contributing to the two blue peaks at 410 and 470 nm. While the 470 nm is 
found to vary with ZnO grain size and their density, the 410 nm peak is 
attributed to the volume of shell surrounding ZnO nanoparticles. This study 
shows the contribution of ZnO cluster size and their density, as influenced 
by film thickness, in the PL spectra of the samples. 
\end{abstract}

\vfil \eject

\section{Introduction}

\par Zinc Oxide (ZnO) is a II-VI wide band gap semiconductor ($E_g=3.2$ eV) that has attracted attention for its light emitting properties \cite{znoEg}. However, due to lack of prominent emission in the red wavelength and 400-500 nm region, attention had been focussed on the fabrication of ZnO:Si nanocomposites with sol gel, rf sputtering and via chemical routes where porous-silicon or nano-silicon is expected to contribute emissions in red region \cite{znoEg}. While these attempts have had some success they present tideous fabrication process. Hence, we adopted simple thermal evaporation method to achieve similar results. Our preliminary analysis \cite{paper1} and thereafter improvement in the properties on annealing \cite{paper2}, showed the potential of thermal evaporation. In the present work, we have investigated the role of film thickness on the photoluminescence emissions. Variation in the grain size, defects within the grains and interfacial regions can be expected with film thickness. Consequently, these changes would get reflected in the photoluminescent property of the ZnO:Si nanocomposite film.   
\par Pelletized mixture of powdered ZnO and n-silicon mixed in a ratio of
1:2 by weight was used as the starting material for fabricating ZnO:Si 
nanocomposites films. The method of fabrication of films has been described elsewhere and therein have explained merits of selecting the 1:2 ratio \cite {paper1}-\cite{paper4}. Thickness of the films were controlled with the help of a quartz digital thickness monitor (DTM-106). All the depositions were made on microscopy glass substrates maintained at room temperature. Samples fabricated with ZnO:Si compositional ratio of 1:2 of thickness 60, 90, 120, 150 and 180 nm are referred to as (c60), (c90), (c120), (c150) in order to identify the samples. The structural studies were done using Philips PW3020 X-Ray Diffractometer (XRD). Photoluminescence (PL) scans were recorded on Fluorolog Jobin Yvon spectroscope (Model 3-11) using an excitation wavelength of 270 nm. Renishaw's ``Invia Reflex'' Raman spectroscope was used for measurements using $\rm Ar^{+2}$ laser.

\section{Results and Discussion}

\begin{figure}[h!!]
\begin{center}
\epsfig{file= 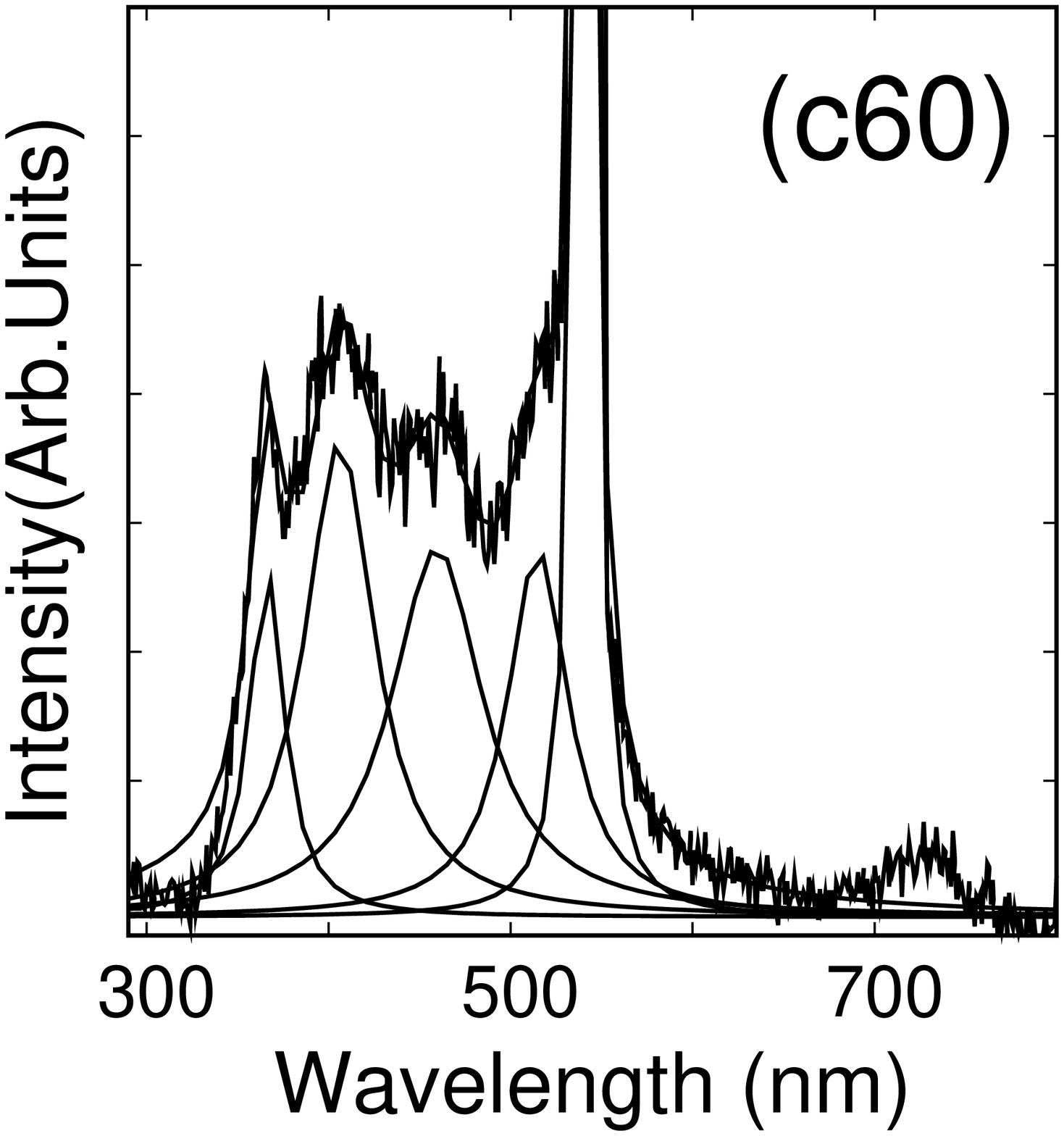, width=1.5in, angle=-90}
\hfil
\epsfig{file= 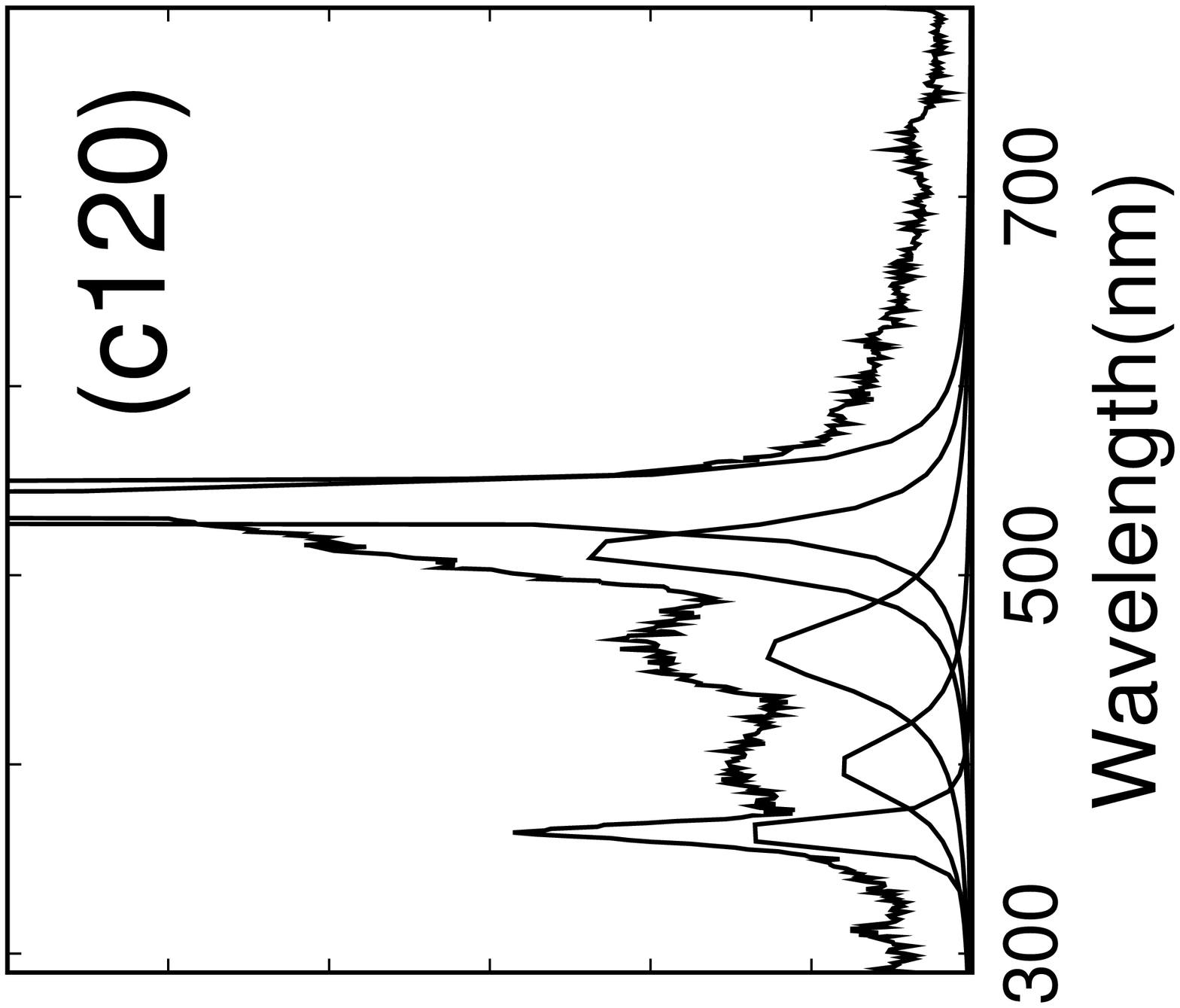, width=1.5in, angle=-90}
\hfil
\epsfig{file= 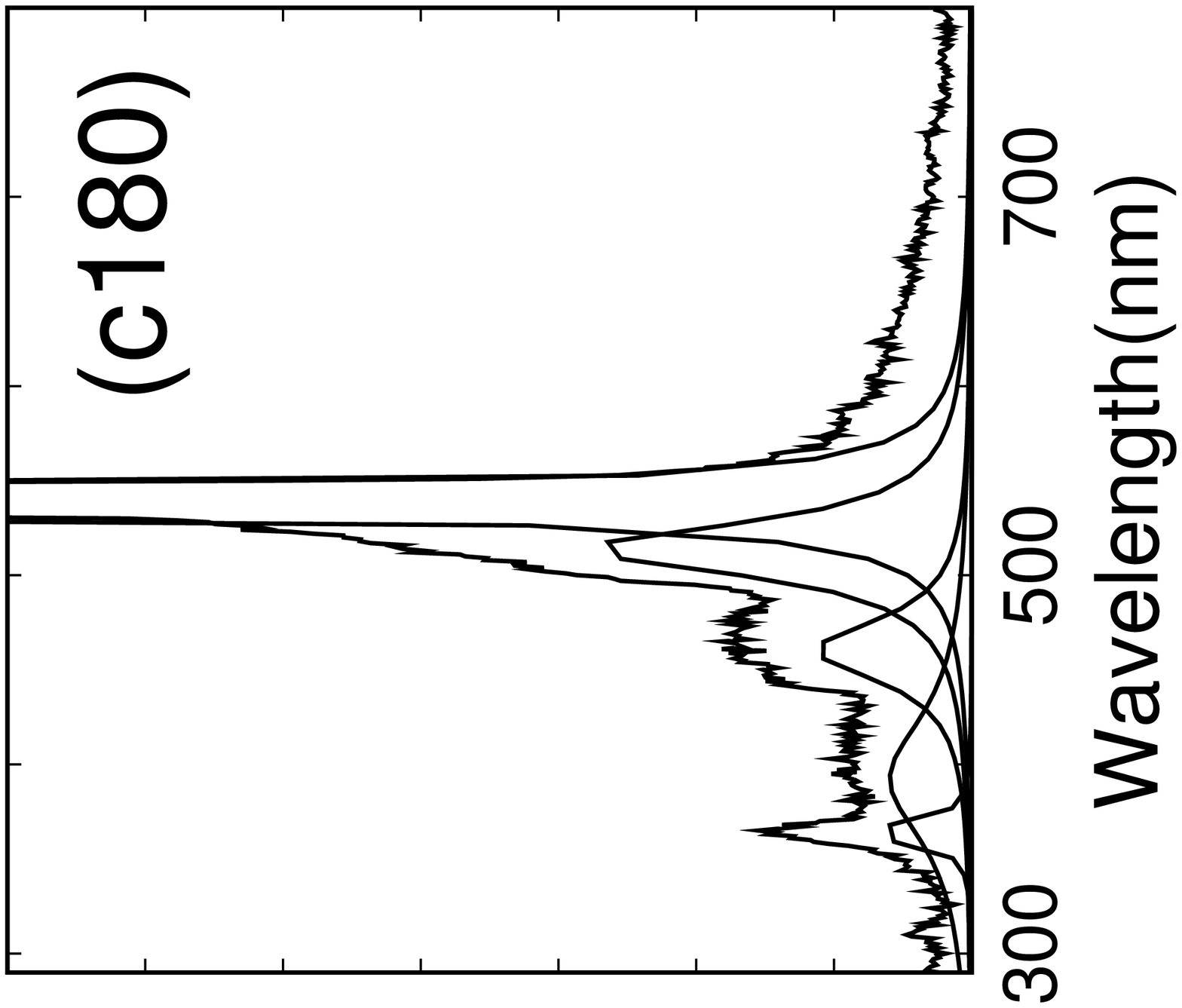, width=1.5in, angle=-90}
\end{center}
\caption{\sl \label{5.1} Representative PL emission spectra showing variation due to film
thickness. 
}
\end{figure}

Figure \ref{5.1} compares the PL behavior of three different thickness nanocomposite films of ZnO:Si
grown with 1:2 mass ratio starting material. A huge second harmonic peak at 540 nm is seen since the 
spectra were recorded without the use of filters. The region between 280-520 nm clearly 
shows variation in PL emission with film thickness, however, all the samples have four common peaks, namely 
at 370, 410, 470, 520 nm. The peaks at 370 and 520~nm are attributed to transitions across the ZnO 
band-gap and between conduction energy level and 
energy level introduced by Oxygen vacancy defects occurring within the ZnO 
grains respectively. However, the peaks at 410 and 470 nm are relatively a
new entrant with recent works attributing it to the 
heterogeneous boundaries \cite{paper2}-\cite{ypeng1}. 

\begin{figure}[h!!]
\begin{center}
\epsfig{file= 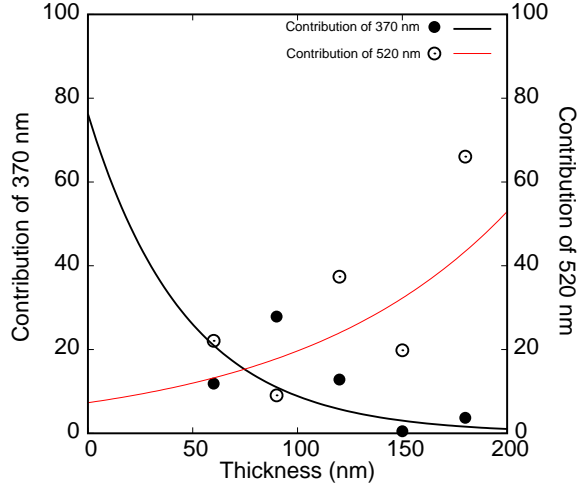, width=2.5in, angle=-90}
\end{center}
\caption{\sl \label{5.2} Variation of percentage contribution of PL emissions of 370 nm  and 520 nm with thickness.}
\end{figure}

The contribution from 370 and 520 nm peaks were found to vary with film thickness (see fig~\ref{5.2}). While the 
contribution of the 370 nm peak decreases with film thickness that from  the 520~ nm shows an exponential increase. 
The nature of curves in fig~\ref{5.2} suggests that the defects increase in ZnO grains with film thickness and 
conversely the perfect wurtzite crystal structure decreases. Earlier works have shown that existence of comparable 
contribution of 370 and 520 nm and inturn amount of lattice with and without defects is a crucial factor to obtain broadening in PL spectra \cite {paper3, paper4}. The region of intersection of the two curves of fig~\ref{5.2}, namely the range 60-90~ nm, marks the sample thickness 
that would potentially give broad emissions (here sample `c60', see fig~\ref{5.1}). Thus the film thickness would play a crucial 
role in the optical properties of these nanocomposite films. We believe that this influence should be manifesting itself as variation 
in ZnO grain size. We now discuss the structural analyses which should throw light on our assumption about role of grain size.


\begin{figure}[h!!]
\begin{center}
\epsfig{file= 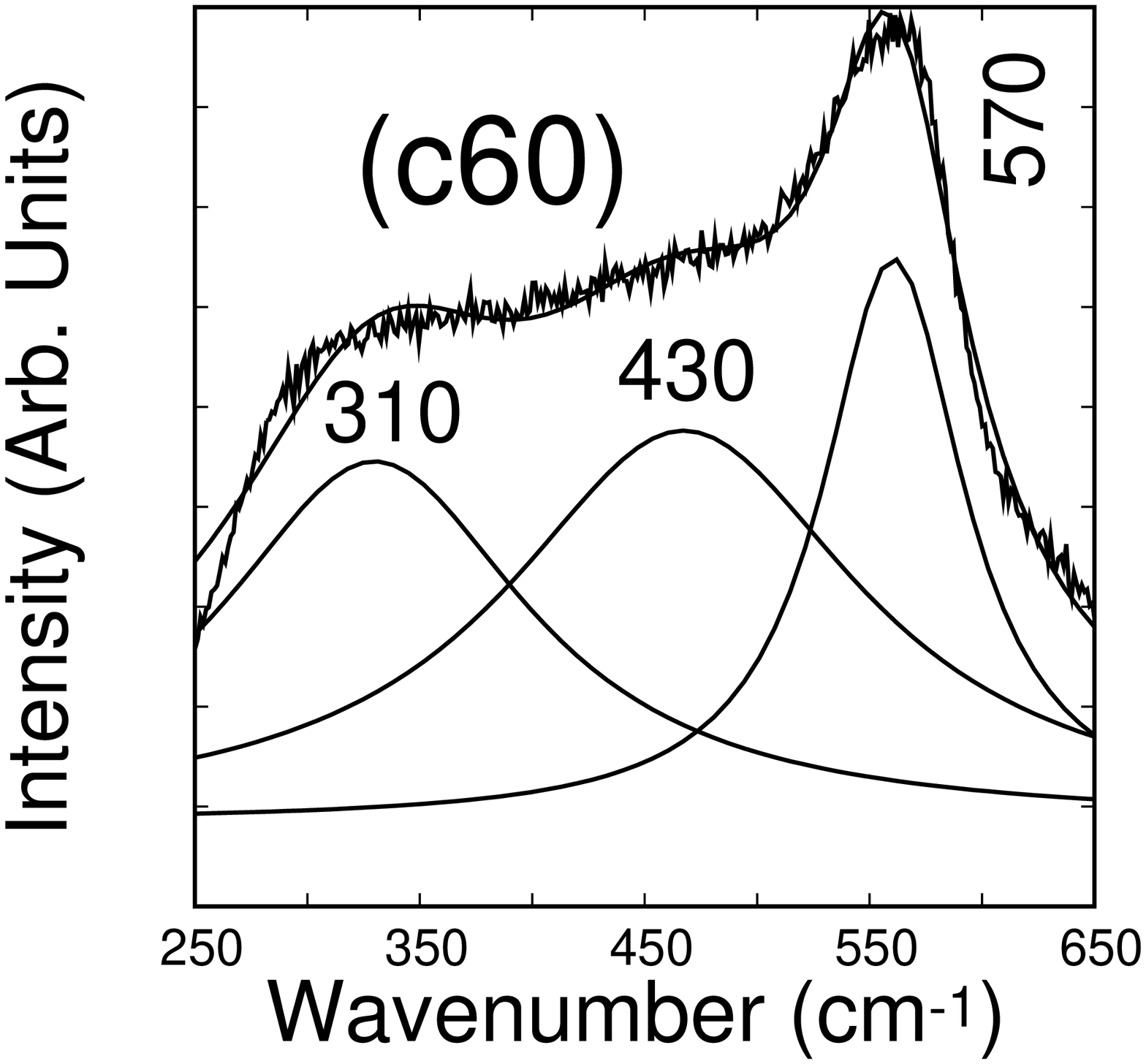, width=1.5in, angle=-90}
\hfil
\epsfig{file= 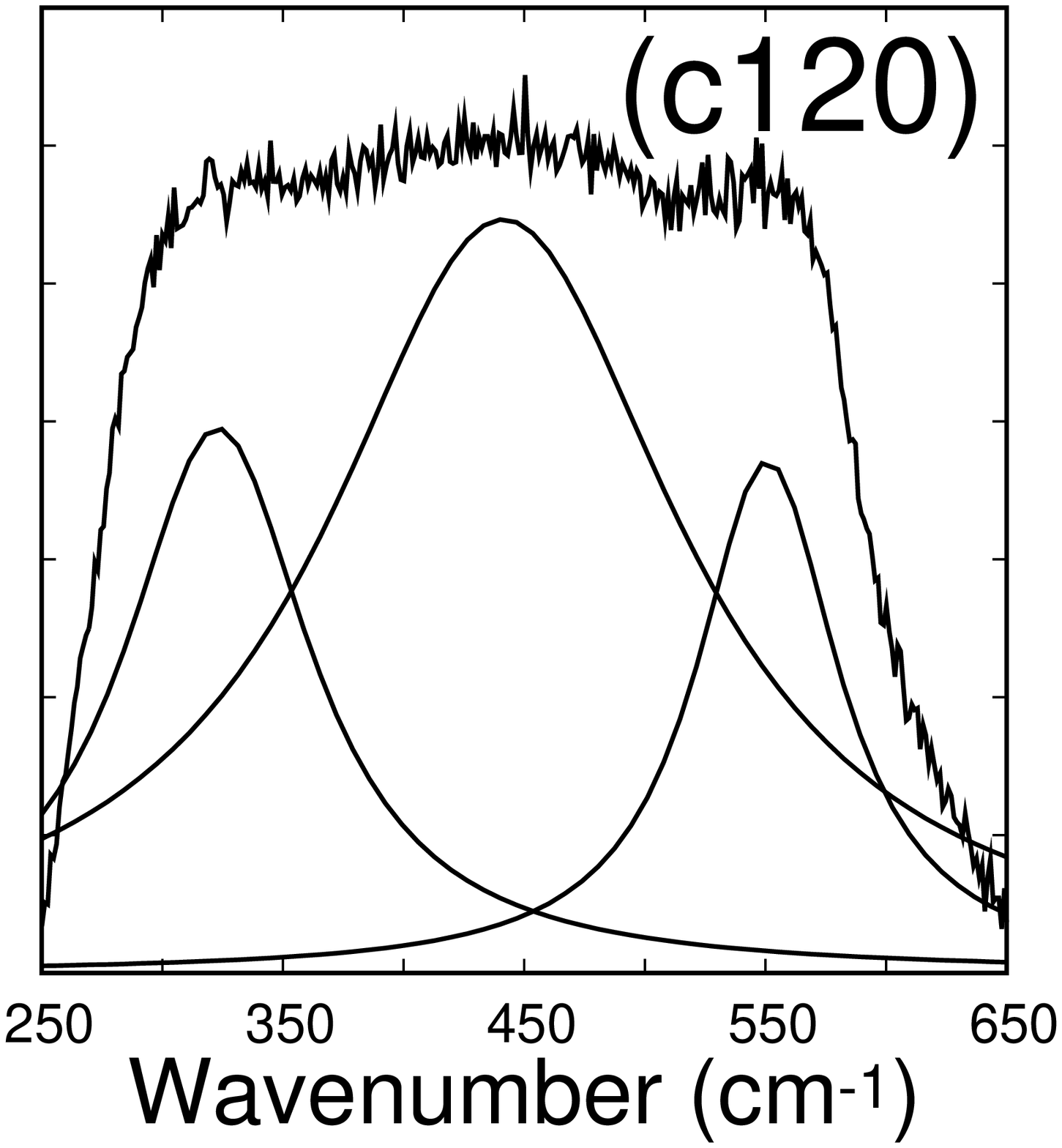, width=1.5in, angle=-90}
\hfil
\epsfig{file= 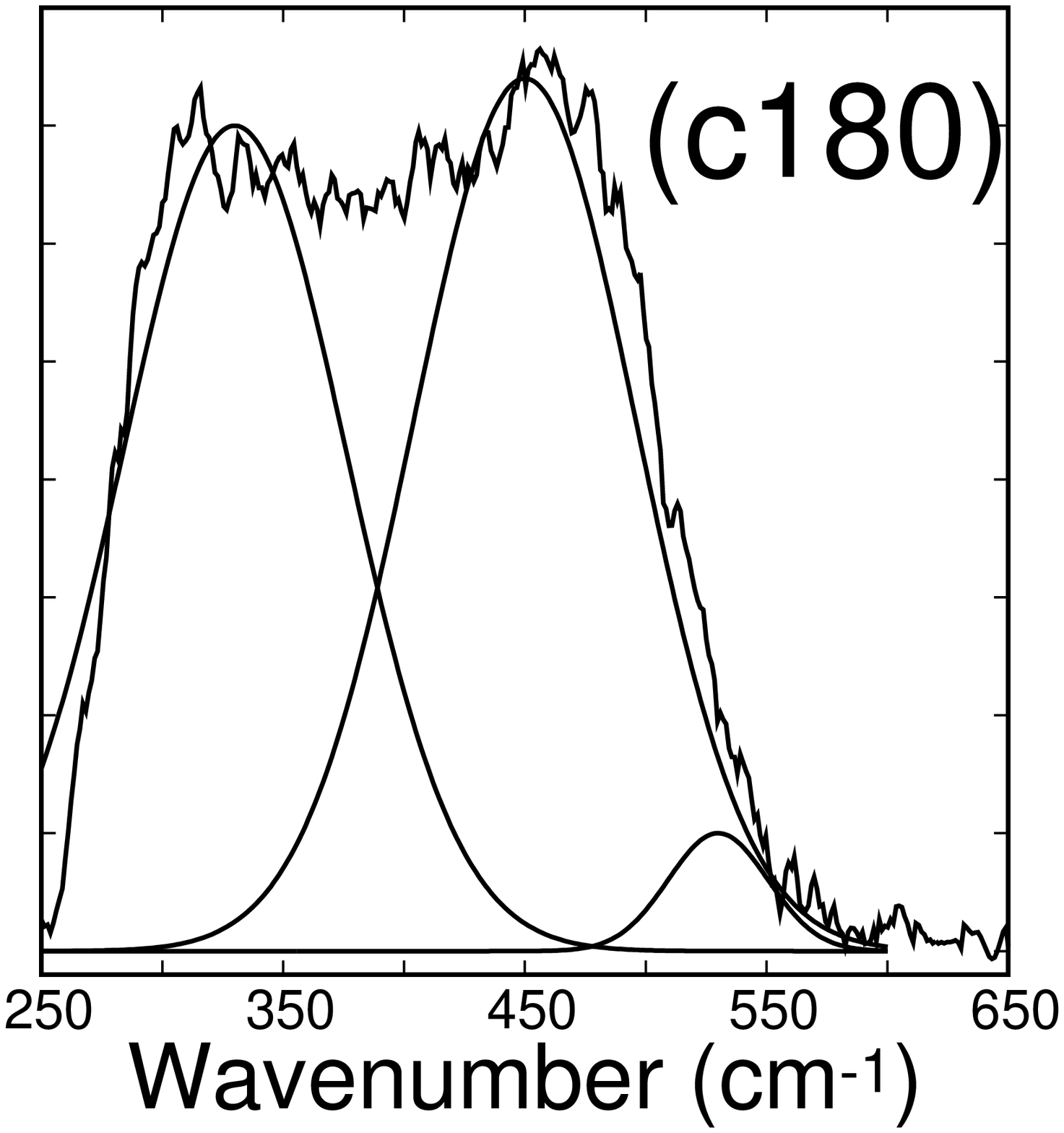, width=1.5in, angle=-90}
\end{center}
\caption{\sl \label{5.3} Raman Spectra of as grown samples `c60', `c120' and `c150'.}
\end{figure}

\begin{figure}[h!!]
\begin{center}
\epsfig{file= 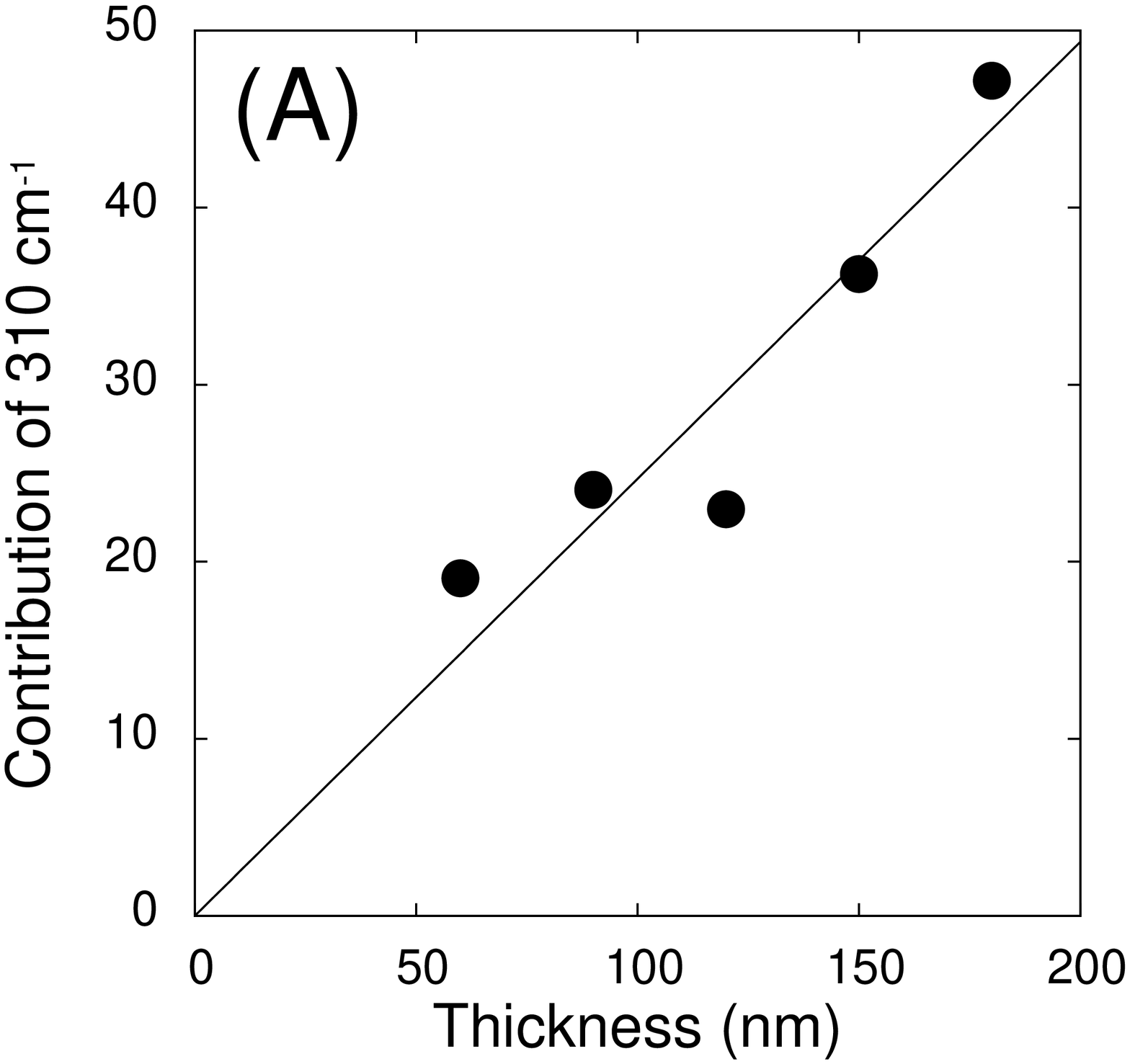, width=1.5in, angle=-90}
\hfil
\epsfig{file= 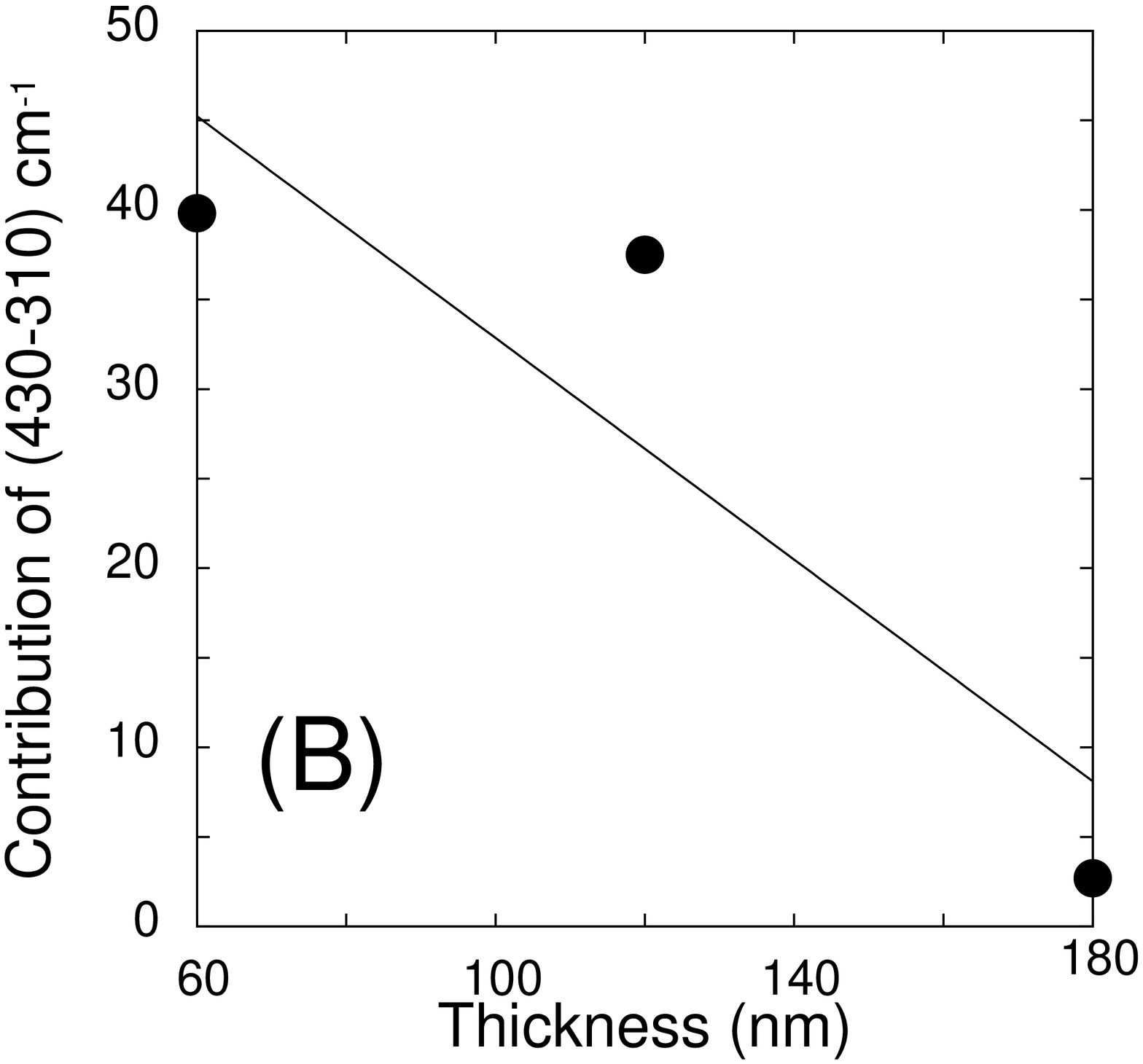, width=1.5in, angle=-90}
\hfil
\epsfig{file= 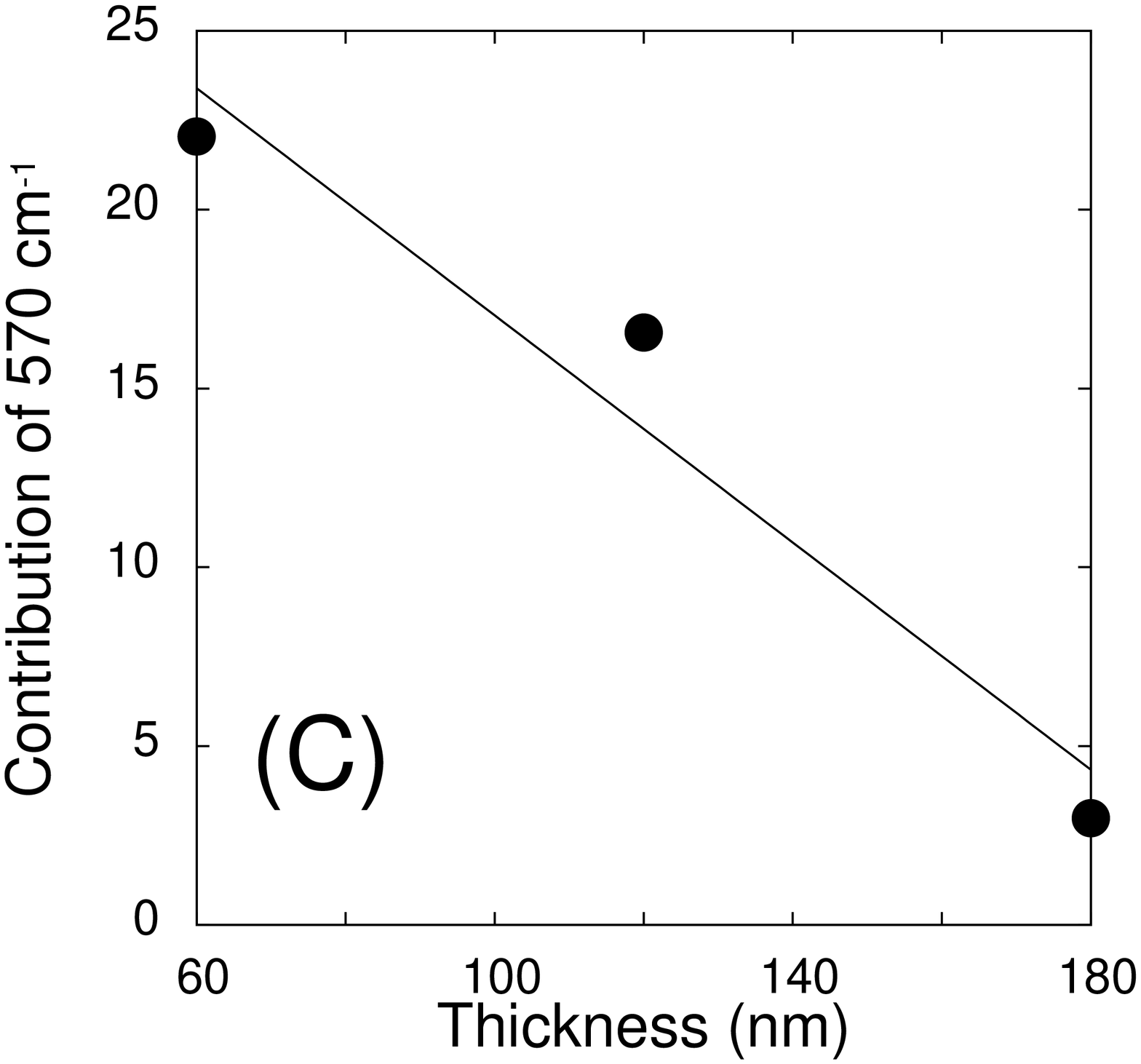, width=1.5in, angle=-90}
\end{center}
\caption{\sl \label{5.4} Variation with film thicknes, the contributions of (A) 310 $\rm cm^{-1}$, (B) wurtzite structured ZnO and (C) Defects related 570 $\rm cm^{-1}$ Raman peak. }
\end{figure}

Fig~\ref{5.3} shows Raman spectra recorded for sample `c60', `c120' and `c150'. Three prominent peaks are seen at 
 310, 430 and 570~ $\rm  cm^{-1}$, consistent with our earlier 
reports \cite{paper1}. While 430 $\rm  cm^{-1}$ is the unresolved peak of 
wurtzite ZnO structure and amorphous silicon, the peak position at 
$\rm \sim $ 570 $\rm cm^{-1}$ corresponds to the defect related mode of ZnO 
grains. The third peak at $\rm \sim $ 310 $\rm cm^{-1}$ is associated with the LA mode of amorphous silicon \cite{paper1, paper2}.
The variation of the 310 ${\rm cm^{-1}}$ Raman peak with film thickness is shown in fig~\ref{5.4}(A). The linear increase in this peak's contribution with thickness indicates that for the given ratio of starting material, the amount of silicon (as compared to ZnO) increases in the thicker films. This would imply `N', the cluster density of ZnO decreases with increasing film thickness. Since the 430 ${\rm cm^{-1}}$ Raman peak is an unresolved peak from amorphous silicon and wurtzite structured ZnO, an estimation of variation in wurtzite structured ZnO with film thickness would require the subtraction of contribution from amorphous silicon (area of 310 ${\rm cm^{-1}}$). The net area of the 430 ${\rm cm^{-1}}$ Raman peak can be thought to be a linear sum being contributed by amorphous silicon and wurtzite and can be expressed as 
\begin{eqnarray}
\Delta_{430}=a\Delta_{ZnO}+b\Delta_{310}
\end{eqnarray}
As a first approximation we can assume a=b=1. Fig~\ref{5.4}(B) and 
(C) shows variation in contribution from wurtzite (peak area 
${\rm \Delta_{430}-\Delta_{310}}$) and defects (${\rm \Delta_{570}}$) with 
thickness. These plots have only three points due to the inability in 
deconvoluting broad peaks due to the lack of prominent shoulders in the 
spectra. The wurtzite contribution decreases with thickness in agreement 
with fig~\ref{5.2} (see fig~\ref{5.4} B). As reported earlier, \cite{paper1, 
paper2}, samples with comparable amount of wurtzite and defects give 
broadening. Since we are comparing samples of different thickness, making 
inferences from the above data is difficult. However relative wurtzite and 
defect presence with film thickness was calculated using the best fit lines 
of Fig~4(B) and 4(C). Figure~5 shows that the ratio of 
${\rm (\Delta_{430}-\Delta_{310})/\Delta_{570}}$ decreases with increasing 
thickness. In short, our Raman data shows that the density (N) of ZnO clusters are decreasing and within them, the lattices with
wurtzite structure diminishes more rapidly than those with defects.
This explains why there is a stronger 520 nm peak in PL as compared to the 370 nm peak in thicker samples. 
\begin{figure}[h]
\begin{center}
\epsfig{file= 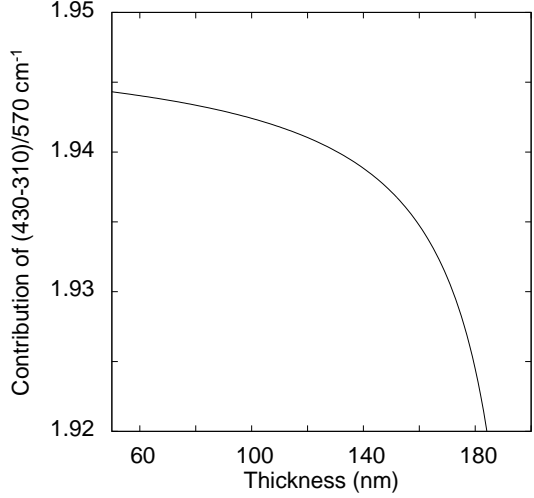, width=2.5in, angle=-90}
\end{center}
\caption{\sl \label{5.10} Relative presence of wurtzite structured ZnO to defects incorporated ZnO (${\rm \Delta_{430}-\Delta_{310}/\Delta_{570}}$) for varying thickness of the film.}
\end{figure}


\begin{figure}[h]
\begin{center}
\epsfig{file= 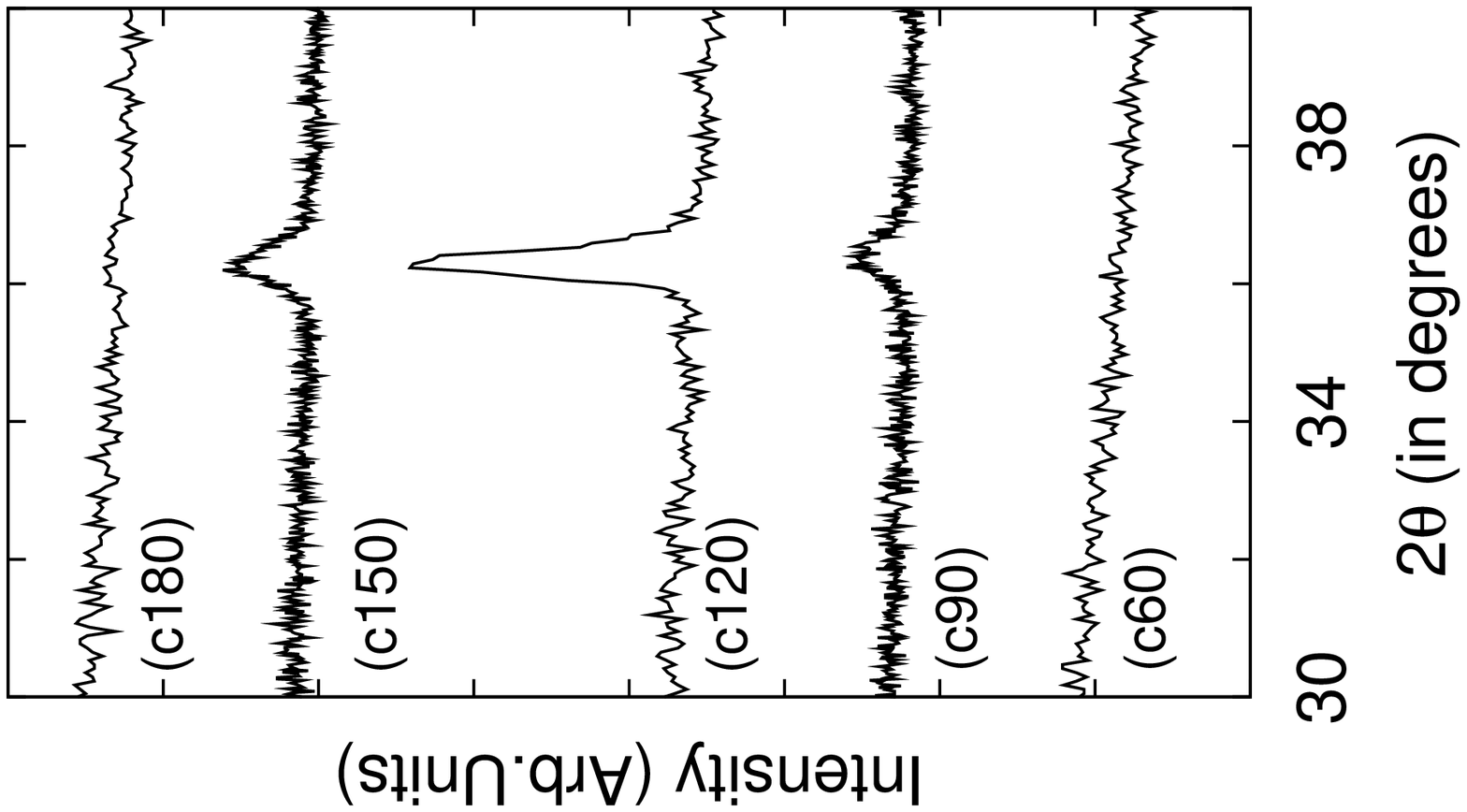, width=3.5in, angle=-90}
\end{center}
\caption{\sl \label{5.5} XRD scans of as grown sample c60, c90, c120, c150 and
c180.}
\end{figure}

For thicker films, X-ray diffractograms exhibit a lone peak at ${\rm 2\theta=36^o}$ that too only in samples `c90-c150' (Fig~\ref{5.5}). This peak is associated with the (101) peak of ZnO (ASTM Card No. 36-1451). Interestingly the Zinc Oxide peak is
absent in `c60' and `c180'. The absence of peak in `c60' has been consistently observed in very thin samples of our studies \cite{paper1, paper2}. Contribution of amorphous silicon is seen to increase with thickness from fig~4(A), implying `c180' films though thicker have low grain density of ZnO, and hence possibly explains the lack of strong diffraction peaks.

\par To understand how film thickness contributes and manifests itself on the optical properties of our sample we calculate the size of ZnO grains from the ${\rm 36^o}$ XRD peak. Calculations are done using standard 
Debye-Scherrer formula given by eqn~
\begin{eqnarray}
g={0.9\lambda \over \beta cos\theta} \label{eq2}
\end{eqnarray}
where $ `\beta$' is full width at half the maximum intensity (FWHM) of the peak \cite{cullity}. The grain size was found to increase with film thickness (Table~I). Viewed in conjucture with results of Raman
spectra we may state that with increasing film thickness, the
population of ZnO grains decreases while their size increases. The volume of
these larger grains enclose higher number of ZnO lattices with defects as compared to those with perfect wurtzite structure.

\begin{table} [h!]
\caption { \sl {Variation of ZnO grain size with thickness of the film.}} 
\vskip 0.25cm
\begin{center}
\begin{tabular}{||c|c||} 
\hline\hline
{\bf Thickness (in nm)} & {\bf Grain Size (in nm) }\\ \hline
{\bf 90} & 7   \\
{\bf 120} &  18  \\
{\bf 150} & 11 \\

\hline\hline
\end{tabular}
\end{center}  
\end{table} 

\begin{figure}[h!!!]
\begin{center}
\epsfig{file= 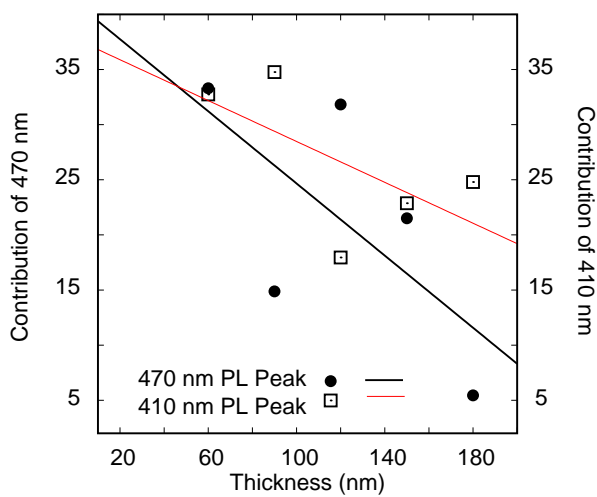, width=2.5in, angle=-90}
\end{center}
\caption{\sl \label {5.7} Variation of percentage contribution of 470 and 410 nm PL peaks with thickness.}
\end{figure}

We now discuss the two (PL) peaks observed in the blue region, at 410 and 
470~nm. In our previous reports, we had shown a linear trend between $\rm \Delta_{470}$ (area enclosed by the 470 nm peak) and $\rm R^2N $ \cite{paper2}. The grain density `N' in our previous study was found to be proportional to the amount of ZnO present in the starting material. However, here the compositional ratio of the starting
material is fixed and hence variation in `N' is co-related to film thickness.
Basically, the variation in contribution of the 470~ nm PL peak ($\rm {\Delta_{470}}$) can be seen as a mathematical problem, with it having a functional dependence on grain size (R) and number of grains (N). In terms of partial differential equation, we write
\begin{eqnarray}
d\Delta &=& {\partial \Delta \over \partial R}
{\partial R \over \partial \tau}d\tau+
{\partial \Delta \over \partial N}
{\partial N \over \partial \tau}d\tau\nonumber\\
{d\Delta \over d\tau} &=& {\partial \Delta \over \partial R}
{\partial R \over \partial \tau}+
{\partial \Delta \over \partial N}
{\partial N \over \partial \tau}\nonumber
\end{eqnarray}
where ${\rm \tau}$ is the film thickness. Table~I and fig~\ref{5.4}(A \& B) 
show that ${\rm {\partial R \over \partial \tau}}$ and 
${\rm {\partial N \over \partial \tau}}$ have opposite signs with 
${\rm {\partial N \over \partial
\tau}}$ being negative. A plot of variation in 470~ nm PL peak contribution 
with respect to thickness (fig~\ref{5.7}, shows a decreasing trend, suggesting 
the dominant influence of `N' over ${\rm R^2 }$.

\begin{figure}[h!!]
\begin{center}
\epsfig{file= 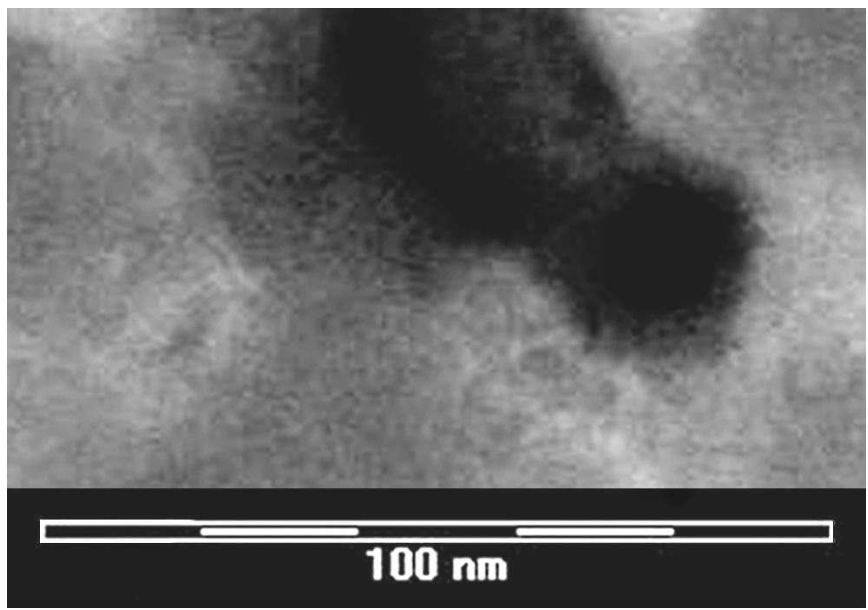, width=4.5in, angle=0}
\end{center}
\caption{\sl \label {5.8} TEM image of sample `c150', clearly shows the formation of a shell encircling ZnO core.}
\end{figure}

 Figure~\ref{5.7} shows the variation in area of 470 
and 410~ nm PL peaks with film thickness. Both peaks show a linear
relationship with the contribution of 470 nm  and 410 nm peaks decrease with 
increasing thickness. Though the trends are similar the physics of the 410 nm 
PL peak is different from that of the 470 nm PL peak. The 410 nm emission is 
believed to result on account of a shell (it's volume) formed by Zn-Si-O 
linkage bonds around the ZnO core \cite{paper4, ypeng1}. The more there are 
grains, the larger is the contribution from the shells. Consequently the 
410 nm is predominantly co-related to the grain density. However, a weak 
influence on the size of the grains can be expected. Figure \ref{5.8} shows 
the TEM micrograph of `c150' which show a shell encircling dense ZnO core embedded in amorphous background. The shell region around a dense core is visible. Core size as calculated from TEM and XRD are fairly in good agreement with each other. While we are able to identify the contributors of emission in blue region, the mechanism of emission is still speculated \cite{willemite2} and needs more investigation.


Revisiting the graph shown in figure \ref{5.7} suggests that the comparable contributions of the two blue peaks can be achieved at the intersection of the best fit line for the two peaks. The intersection point is observed at a film thickness of $\sim$ 60 nm. Interestingly, the relative contribution of 370 nm and 520 nm peaks of PL also intersect
at a thickness (fig ~\ref{5.2}) of $\sim$ 90 nm. 

\par From the previous section it became clear that the broadening effect of 370 
and 520 nm of PL is largely related to the nature of ZnO grains namely 
relative presence of lattice with defects and those with perfect wurtzite 
structure. These two peaks typically do not show dependance on grain size, 
however our analysis suggests for the conditions in which our samples were fabricated, thicker films have lower grain density abid 
larger grains with and more defects. Thus, larger grains are more prone to 
contain defects, obliquely introducing a grain size dependancy.          

\par The 410 and 470 nm PL peaks similarly show grain size dependancy based on the idea that smaller grains give larger grain population (N). With better control over film growth parameters, an optimum condition could be obtained. For the condition of film fabrication in this study we find best results are obtained for 60 nm thick films. While the density of grains (N) in a fabricated film cannot be changed due to the clusters being embedded in an isolating matrix, the shell and the defect concentration can be altered by post deposition annealing, as suggested in our recent work \cite{paper4}.

\section*{Conclusion}
In summary, the PL peaks at 370, 520, 410 and 470 nm are correlated to the thickness of the thermally evaporated ZnO:Si nanocomposite films.
 We are successful in explaining the origin of the photoluminescence peaks at 410 and 470 nm. Though both are attributed to interface occurring at heterogeneous boundary of ZnO nanoclusters and amorphous silicon, they are different in the physics lying behind their origin. It was found that both the peaks are related to number of clusters (N) and the size of grains (R), which are strongly influenced by the thickness of the films. The investigations indicate that whereas 410 nm comes from Zn-Si-O linkages existing in a shell around the ZnO clusters, 470 nm varies linearly with $\rm R^2N$. Findings of the present work thus suggests that at an optimum thickness, the four PL peaks in discussion would lead to equal intensity broad-band luminescence. For the thermally grown ZnO:Si nanocomposites, the PL peaks were found to merge at a thickness of $\rm \sim$ 60 nm. We are hopeful that this study would serve as a way to monitor the interfacial region existing in nanocomposites in general and particularly be useful in tailoring the luminescence properties of ZnO:Si nanocomposites.

\section*{Acknowledgment}
The resources utilized at University Information Resource Center, Guru Gobind Singh Indraprastha University  and Department of Geology, University of Delhi is gratefully acknowledged. We also would like to express our gratitude to Dr. Kamla Sanan and Dr. Ritu Srivastava at National Physical Lab., Delhi for carrying out the photoluminescence studies. Author CRK is thankful to University Grants Commission (India) for financial assistance  vide F.No-33-27/2007(SR). We acknowledge the infrastructure used under the sanction from Department of Science and Technology (India) SR/NM/NS-28/2010. One of the authors (Shabnam) is grateful to Council of Scientific and Industrial Research (India) for providing Senior Research Fellowship.


\vfil \eject
\pagestyle{empty}
\newpage


\begin{thebibliography}{99}
\bibitem{znoEg} C. Jagadish and S. J. Pearton, {\sl ``Zinc Oxide Bulk, Thin
Films and Nanostructures, Elsevier Ltd.''}, 2006.
\bibitem{paper1} S.Siddiqui, C.R.Kant, P.Arun and N.C.Mehra. Phys. Lett. A {\bf372} (2008) 7068.

\bibitem{paper2}Shabnam, C.R.Kant and P.Arun. Mater. Res. Bull. {\bf 45} 
(2010) 13068.
\bibitem{paper3} Shabnam, C.R.Kant and P.Arun. Size and Defect related Broadening of Photoluminescence Spectra in ZnO:Si Nanocomposite Films. {\sl communicated} available at arXiv:1007.2142.
\bibitem{paper4} Shabnam, C.R.Kant and P.Arun. White Light emission from annealed ZnO:Si Nanocomposites Thin Films. {\sl communicated} available at arXiv:1101.4119.
\bibitem{ypeng1} Yu-Yun Peng, Tsung-Eong Hseih and Chia-Hung Hsu, Nanotechnology, {\bf 17} (2006) 174.
\bibitem{cullity} `` Elements of X-Ray Diffraction'', B.D.Cullity (London,1959).
\bibitem{willemite2} X.Feng, X.Yuan, T.Sekiguchi, W.Lin and J.Kang, J.Phys.Chem. B, {\bf 109} (2005) 15786.













\end {thebibliography}
\end{document}